# Ultrahigh-speed digital holography for quantitative Doppler imaging of the human retina

**Zacharie Auray, Yann Fischer, Olivier Martinache, Michael Atlan**

*Institut Langevin. Centre national de la recherche scientifique (CNRS) UMR 7587. Paris Sciences et Lettres (PSL) University, Ecole Supérieure de Physique et de Chimie Industrielles ESPCI Paris. 1 rue Jussieu. 75005 Paris. France.*
*Author e-mail address: zauray@gmail.com*

**Abstract:** High-speed digital holography reveals local blood flow contrasts in the eye fundus through deterministic signal analysis, leveraging a forward scattering model of dynamically diffused light. This approach enables the estimation of absolute blood flow in primary in-plane retinal arteries. © 2023 The Author(s)

## 1. Introduction

High-speed digital holography, a method for imaging hemodynamics in ophthalmology [1], has been enhanced to quantitatively measure retinal blood flow during the cardiac cycle using a forward scattering model of diffuse dynamic light from deeper retinal layers. This imaging technology enables non-invasive measurement of hemodynamic parameters in the retina. It enables precise, quantitative assessments of blood perfusion rates and arterial hemodynamic resistivity, providing unprecedented insights into ocular vascular dynamics.

## 2. Doppler holography for ophthalmology. Clinical device prototype.

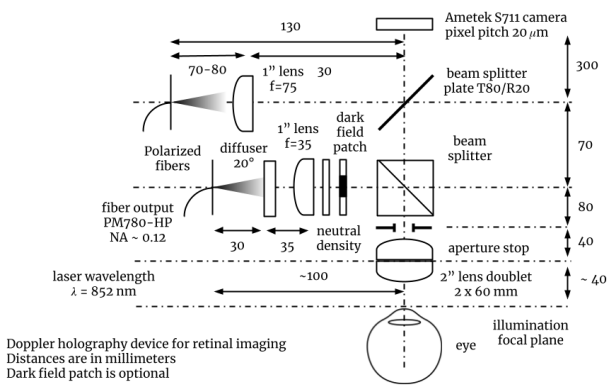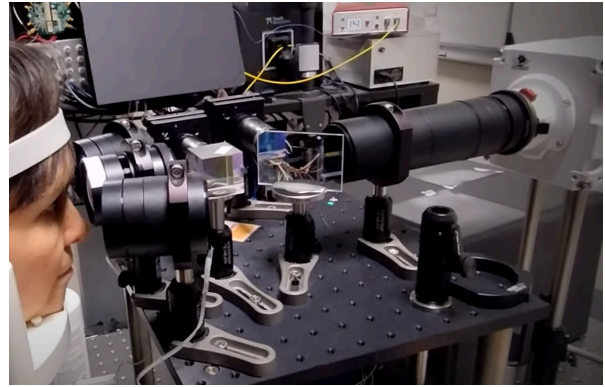

Figure 1: Real-time, clinical-grade Doppler holography prototype. Optical setup. An inline Mach-Zehnder interferometer operating in the near-infrared range combines light backscattered from the patient's eye fundus with a reference beam. The resulting optical interference patterns are captured in real time by a streaming camera.

The experimental setup utilizes a Mach-Zehnder inline interferometer (Fig. 1). Near-infrared radiation from a diode laser (Thorlabs FPV852P, wavelength: $\lambda$ = 852 nm) is split into linearly-polarized reference (10%) and illumination (90%) arms, using polarization-maintaining fibers (Thorlabs PM780-HP, numerical aperture: NA ~ 0.12). The illumination beam is diffused through an engineered diffuser (Thorlabs ED1-C20-MD, 1" diameter, 20° circle tophat) and focused by an eyepiece composed of two biconvex lenses with a combined effective focal length of ~33 mm. A real-time camera (Ametek Phantom S711 with Euresys Coaxlink QSFP+ frame grabbers, pixel pitch: 20 µm, operated at a maximum frame rate of 37 kHz) captures the images. The retina of a volunteer is illuminated by the diffuse laser beam focused through the eyepiece. The cross-polarized backscattered light interacts with the reference beam, creating interferogram patterns that are captured by the high-speed camera.

## 3. Image acquisition

A high-speed, wide-field Doppler holography (using an 852 nm near-infrared laser) can safely and noninvasively image endoluminal blood flow in the main retinal arteries and veins under diffuse illumination in a Maxwellian view, meeting U.S. and European safety standards without compromising image quality.

## 4. Image rendering

Real-time rendering of digital holograms from high-throughput interferogram streams is done with the open-source software Holovibes [2]. It integrates advanced spatial and temporal demodulation techniques, including the Fresnel transform, angular spectrum propagation, short-time Fourier transform, and principal component analysis. Optimized with CUDA and GPU programming, it enables high-speed, low-latency holographic imaging. Holovibes achieves holographic image rendering and analysis at 37,000 frames per second for $512 \times 320$-pixel images on standard hardware, without frame loss. It supports the simultaneous recording of raw and processed images, making 6 Gpixels/s computational imaging a practical clinical tool. By leveraging GPUs and ultra-high-speed cameras, The reliability of this software facilitates advanced clinical imaging applications.

## 5. Quantitative analysis

Utilizing near-infrared laser light, this technique enables non-invasive angiographic imaging of retinal and choroidal vessels, providing detailed contrast of blood flow and revealing dynamic variations throughout the cardiac cycles. The quantitative estimation of total retinal blood flow using Doppler holography is achieved through a simple and robust process that leverages the optical and topological properties of the retina. The proposed method consists of:

1. Segmentation and Measurement: The main retinal arteries in the imaging plane are segmented. The local blood velocity is calculated by measuring Doppler frequency broadening relative to an interpolated background signal from surrounding tissues. If the local signal does not precisely reflect the local background of the retinal vessels, it may lead to underestimation or overestimation of the local differential Doppler broadening. These variations, including the sign of the frequency difference, are preserved, and statistically, the variability cancels out across the entire field of view.

2. Estimation of Local Blood Velocity: A forward light diffusion model integrates a secondary light source scattered from the sclera, passing through the choroid and deep retinal layers. The blood flow velocity in retinal arteries is derived from local differential Doppler broadening, requiring only the optical wavelength and the numerical aperture of the eye for an accurate velocity estimation.

3. Blood Flow Calculation: Doppler frequency broadening profiles are fitted to a Poiseuille flow model to determine the endoluminal cross-sectional area of the retinal arteries with micrometric precision. The local blood velocity is then multiplied by the cross-sectional area to compute the absolute volumetric flow rate (Fig. 2). The stability of blood flow estimates in the main retinal arteries over several millimeters from the optic disc provides initial evidence that retinal blood flow is conserved between proximal and distal measurements. These estimates also exhibit high repeatability, highlighting the reliability of this approach. These findings suggest that this technology has significant potential for routine quantitative assessment of total and local retinal blood flow dynamics, making it a valuable tool for disease diagnosis and treatment monitoring.

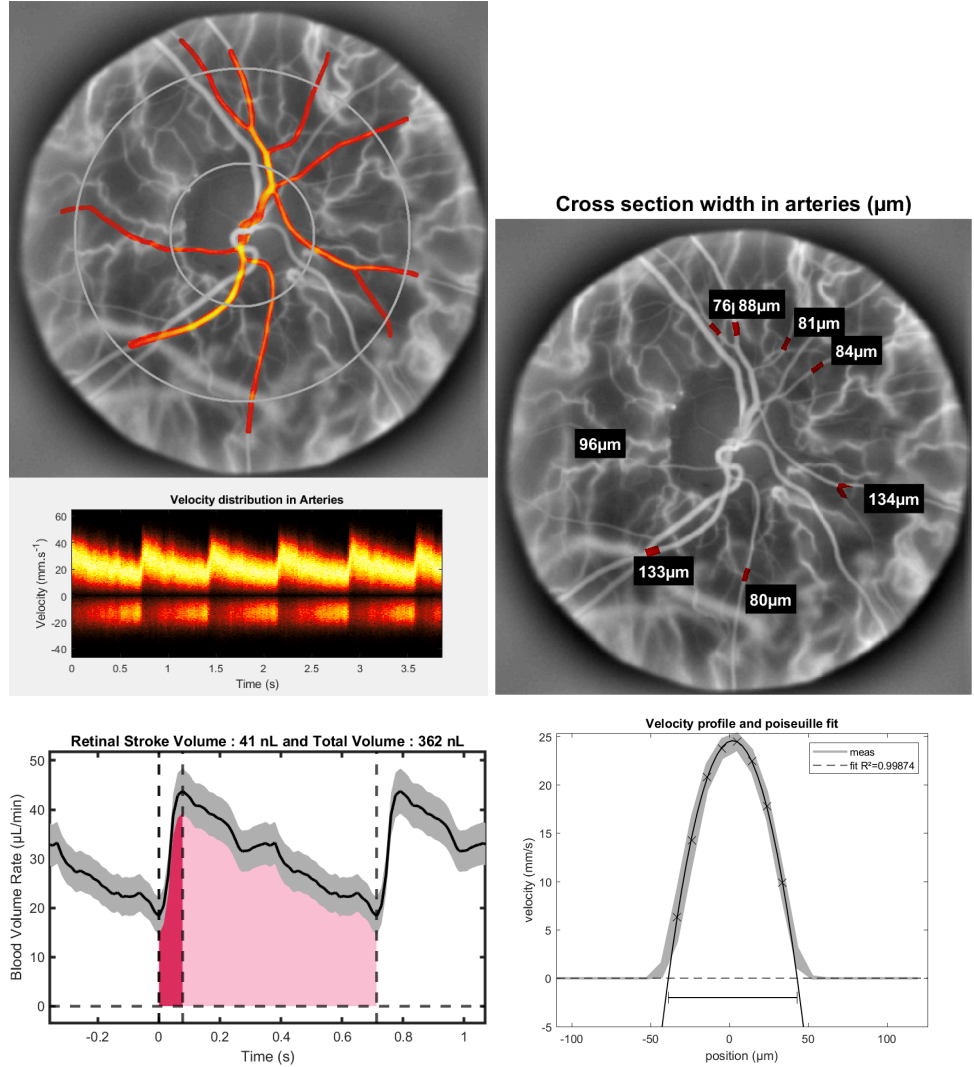

Figure 2: Quantitative measurement of retinal blood flow using real-time high-speed Doppler holography (6 Gpixels/s), feasible for routine use with the prototype. Data acquisition is performed by holovibes [2], while analysis is fully automated using holoDoppler [3] and pulsewave [4], without human intervention.

## 6. Conclusion

Real-time Doppler holography is an open-source technology with the potential to revolutionize retinal vascular monitoring for the most prevalent eye pathologies including glaucoma an diabetic retinopathy. By providing precise, quantitative measurements of retinal blood flow, it enables the detection of subtle hemodynamic changes during treatments. Beyond ophthalmology, its applications extend to cardiology, offering valuable insights into systemic cardiovascular conditions, including heart failure, atherosclerosis, stroke, and hypertension. The system is designed for efficiency and ease of use, featuring an intuitive workflow with one-click video acquisition and automated analysis via open-source software. This ensures high-throughput, repeatable screenings at scale.